\documentclass[runningheads]{llncs}
\usepackage[T1]{fontenc}
\usepackage{graphicx,verbatim}
\usepackage{amsmath}
\usepackage{amssymb}
\usepackage{bm}

\usepackage{booktabs}
\usepackage{makecell}
\usepackage{multirow}
\usepackage{siunitx}
\usepackage{threeparttable}

\usepackage{color}

\usepackage[hidelinks]{hyperref}

\begin{document}
\title{Uncertainty Quantification in Cardiac Model Personalisation from Ultrafast Ultrasound}
\titlerunning{Uncertainty Quantification in Cardiac Model Personalisation}
%

\author{%
Camilla Ferrario\inst{1,3} \and
Maelys Venet\inst{2} \and
Olivier Villemain\inst{2} \and
Maxime Sermesant\inst{1,3}%
}

\authorrunning{C. Ferrario et al.}

\institute{%
 Université Côte d'Azur, Inria, Epione Team, Sophia Antipolis, France
\and
Bordeaux University Hospital (CHU de Bordeaux), Bordeaux, France
\and
IHU Liryc, Electrophysiology and Heart Modeling Institute, Pessac, France\\
\email{camilla.ferrario@inria.fr}
}

\maketitle             
\begin{abstract}
Cardiac model personalisation requires inferring mechanical parameters that are not directly measurable \textit{in vivo}. Ultrafast ultrasound shear wave elastography (SWE) enables non-invasive tracking of myocardial stiffness dynamics over the cardiac cycle, providing a target for personalisation. However, mapping these observations to subject-specific model parameters remains ill-posed, as multiple parameter sets can reproduce the same stiffness dynamics. We formulate SWE-informed personalisation as a statistical inference problem using simulation-based inference (SBI). Using a subject-adapted \(0\)D cardiovascular model and neural posterior estimation, we estimate model-conditional posterior distributions over active stiffness scale \(k_0\), contraction rate \(k_{\mathrm{ATP}}\), and relaxation rate \(k_{\mathrm{SR}}\), conditioned on SWE-derived curve features and subject-specific context. Among six healthy volunteers, four passed objective prior-support diagnostics and were retained for quantitative posterior analysis. Curve-level RMSE against the observed SWE target decreased from \(12.61\pm5.55\,\mathrm{kPa}\) for the prior predictive median to \(1.14\pm0.38\,\mathrm{kPa}\) for the posterior predictive median, an \(89.7\pm4.2\%\) reduction. Posterior analysis revealed parameter-specific uncertainty, \(k_0\)-\(k_{\mathrm{ATP}}\) compensation, weaker constraint of \(k_{\mathrm{SR}}\), and the importance of prior-predictive diagnostics for assessing whether each subject is represented within the modelled SWE feature space. These results support SBI for uncertainty-aware SWE-based personalisation, while identifying prior support and forward-model adequacy as key diagnostics.

\end{abstract}

%

\section{Introduction}

Personalised cardiac models support the interpretation of clinical observations by linking measured cardiac signals to subject-specific mechanical properties~\cite{SermesantEtAl2012,ChabiniokEtAl2016}. Model personalisation is usually posed as an inverse problem, where parameters are calibrated from limited measurements such as imaging-derived motion, deformation, ventricular volumes, or pressure points~\cite{SermesantEtAl2008,BracamonteEtAl2022,BanusEtAl2021}. Clinical data are often sparse, and cardiac model parameters may interact. Consequently, similar observations may be explained by different parameter combinations, leading to an ill-posed personalisation problem with uncertain, correlated, or only partially identifiable parameter estimates~\cite{VillaverdeEtAl2019,MolleroEtAl2019}. Reliable personalisation should therefore go beyond a single best-fit parameter vector and quantify which properties are actually constrained by the available measurements~\cite{ChabiniokEtAl2016,VillaverdeEtAl2019,SalvadorEtAl2023a}.

These limitations are particularly relevant when personalising myocardial mechanics, where passive stiffness, active contraction, and relaxation are inferred from global measures of pump function, such as ejection fraction or pressure-volume loops~\cite{VillalobosLizardiEtAl2022}. Because these measurements are strongly influenced by loading conditions and circulatory coupling, they constrain tissue properties only indirectly. Ultrafast shear wave elastography (SWE) offers a more direct, tissue-level measure of myocardial stiffness throughout the cardiac cycle~\cite{PernotEtAl2011,VillemainEtAl2020}. Nevertheless, SWE-derived stiffness combines passive tissue response, active force generation, and measurement variability. Fitting stiffness dynamics alone therefore cannot identify a unique set of active mechanical parameters or determine which parameters are effectively constrained by the data.

These challenges motivate a probabilistic formulation that estimates the range of parameter values compatible with the observed SWE dynamics and subject-specific context. Conventional Bayesian approaches such as Markov chain Monte Carlo typically require repeated likelihood evaluations, while approximate Bayesian computation depends on user-defined distance functions and acceptance thresholds. Simulation-based inference (SBI) instead learns an approximation of the posterior directly from simulated data, making it well suited to complex models with intractable likelihoods~\cite{CranmerEtAl2020,DeistlerEtAl2025}. Neural SBI can additionally amortise inference across observations while capturing non-Gaussian uncertainty and dependencies between parameters. In cardiovascular modelling, however, SBI has so far been applied primarily to biomarker estimation from haemodynamic biosignals~\cite{WehenkelEtAl2024,ManduchiEtAl2024a}.

Here, we introduce a context-conditioned SBI framework for SWE-informed personalisation of active myocardial mechanics. We combine a subject-adapted \(0\)D cardiovascular model with neural posterior estimation to infer distributions of active mechanical parameters from stiffness-curve features and subject-specific context. The main contributions are: (i) a probabilistic formulation of SWE-based active-parameter estimation; (ii) a feature-based observation model that accounts for uncertainty in the stiffness targets; and (iii) a preliminary evaluation in healthy volunteers using posterior predictive checks (PPCs) and comparison with an optimisation-based point estimate.


\section{Methods}

\subsection{Ultrafast SWE-derived stiffness target}
\label{sec:swe_target}

The inference target is the subject-specific myocardial stiffness dynamics measured by ultrafast SWE. Data were acquired in a parasternal long-axis view, targeting the basal anteroseptal interventricular septum, using a Verasonics Vantage system with a GE 6S-D probe. Acoustic radiation force pushes were transmitted at $4~\mathrm{MHz}$ for $300~\mu\mathrm{s}$ with an f-number of $1$. Acquisitions were ECG-gated at 20 phases distributed over the cardiac cycle. At each phase, shear-wave velocity estimates within the segmented mid-myocardial region were spatially averaged. The mean shear-wave speed, $c_s$, was converted to apparent Young's modulus, $E$, using $E=3\rho c_s^2$, assuming a tissue density of $\rho=1000~\mathrm{kg/m^3}$. The resulting values were assembled into a cycle-normalized stiffness curve $E_{\mathrm{SWE},d}(s)$, with $s\in[0,1]$, for each analysed subject $d=1,\ldots,D$. This curve was used directly as the observation for inference:
\begin{equation}
E_{\mathrm{obs},d}(s)=E_{\mathrm{SWE},d}(s).
\label{eq:observed_swe_target}
\end{equation}
Acquisition and stiffness reconstruction followed previously established protocols~\cite{VillalobosLizardiEtAl2022,VillemainEtAl2020} and were performed under ethics approval with informed consent.

\subsection{Context-adapted \(0\)D cardiac model}
\label{sec:context_model}

The forward model is a reduced-order \(0\)D cardiovascular simulator. It represents the left ventricle as a thick-walled spherical chamber coupled to a closed-loop systemic RCR and pulmonary RC Windkessel circulation. Under an isotropic fibre assumption, the model reduces ventricular geometry to spherical symmetry while retaining the tissue constitutive description~\cite{CaruelEtAl2014,FerrarioEtAl2025}.

For each case \(d\), we denote by \(\mathbf{c}_d\) the available
subject-specific anatomical, physiological, and demographic context, including
left ventricular end-systolic and end-diastolic volumes (LVESV, LVEDV),
interventricular septal thickness at end diastole (IVSd), heart rate (HR), age,
sex, and body surface area (BSA). HR, IVSd, and LVEDV define the case-specific
cardiac period \(T_d\), reference wall thickness \(d_{0,d}\), and reference
radius \(R_{0,d}\), respectively:
\begin{equation}
    T_d = \frac{60}{\mathrm{HR}_d},\qquad
    d_{0,d} = 10^{-2}\,\mathrm{IVSd}_d,\qquad
    R_{0,d} =
    \left(
    \frac{3 \cdot 10^{-6}\,\mathrm{LVEDV}_d}{4\pi}
    \right)^{1/3},
    \label{eq:context_mapping}
\end{equation}
with HR, IVSd, and LVEDV expressed in \(\mathrm{bpm}\), \(\mathrm{cm}\), and
\(\mathrm{mL}\), respectively. LVESV, age, sex, and BSA enter only as
conditioning variables.

Personalisation is restricted to the active parameter vector
\(\boldsymbol{\theta}=(k_0,k_{\mathrm{ATP}},k_{\mathrm{SR}})\). The parameter
\(k_0\) (\(\mathrm{kPa}\)) controls the active stiffness amplitude,
\(k_{\mathrm{ATP}}\) (\(\mathrm{s}^{-1}\)) the contraction rate, and
\(k_{\mathrm{SR}}\) (\(\mathrm{s}^{-1}\)) the relaxation rate. Passive constitutive, circulatory, and valve parameters are fixed across subjects, while the observation-level passive stiffness scale is calibrated subject-wise before SBI.

For case \(d\), the simulator predicts the active stiffness response
\(E_{\mathrm{act},d}^{\mathrm{sim}}(s;\boldsymbol{\theta})\). Since SWE measures
total apparent myocardial stiffness, the simulated inference target is
\begin{equation}
    E_{\mathrm{sim},d}(s;\boldsymbol{\theta})
    =
    E_{\mathrm{pass},d}(s)
    +
    E_{\mathrm{act},d}^{\mathrm{sim}}(s;\boldsymbol{\theta}),
    \label{eq:total_swe_sim}
\end{equation}
where $E_{\mathrm{pass},d}(s)$ is precomputed for each subject using the
stretch-dependent passive observation model and late-diastolic calibration
detailed in~\cite{FerrarioEtAl2026}, and is subsequently held fixed during SBI.
Training features and PPCs are derived from $E_{\mathrm{sim},d}$ and compared
with $E_{\mathrm{obs},d}$.

\subsection{Context-conditioned SBI}
\label{sec:sbi}

\begin{figure}[ht]
    \centering
    \includegraphics[width=\linewidth]{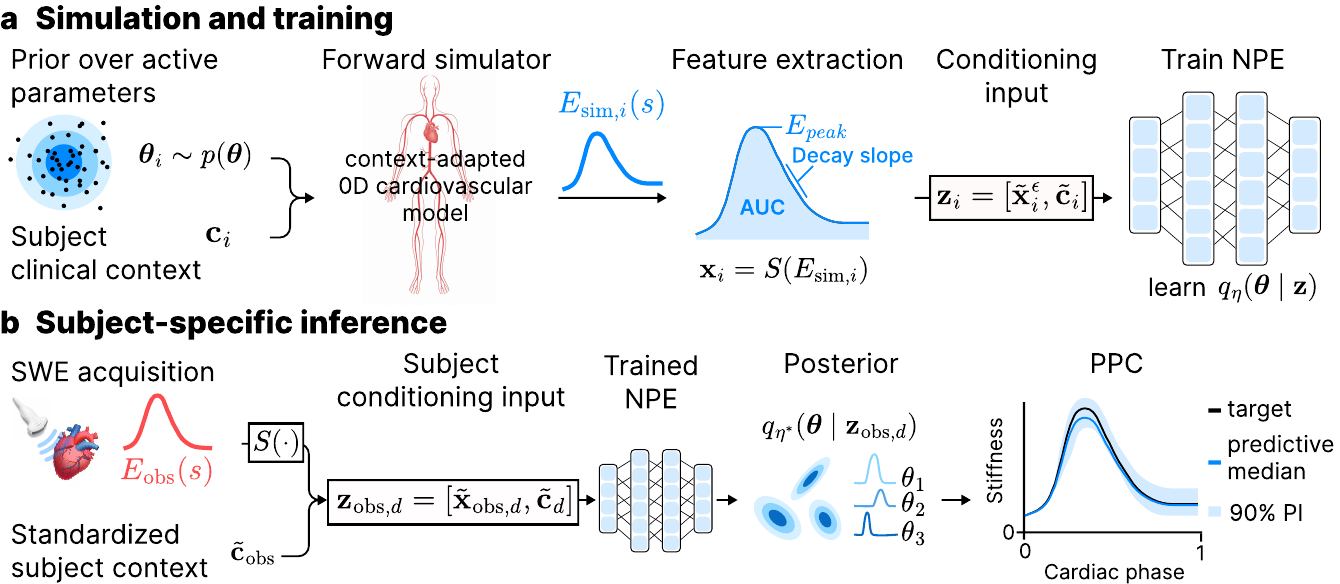}
    \caption{\textbf{SWE-informed SBI workflow.} \textbf{(a)} Prior samples and subject context are passed through the \(0\)D model to generate stiffness curves. Extracted features are noise-augmented, standardized, concatenated with standardized context, and pooled to train a shared NPE \(q_{\eta}(\boldsymbol{\theta}\mid\mathbf{z})\). Here, \(i\) indexes a pooled sample, i.e. one case, parameter draw, and feature-noise augmentation. \textbf{(b)} For case \(d\), observed SWE features and context define \(\mathbf{z}_{\mathrm{obs},d}\). The trained NPE returns a subject-specific posterior, evaluated by PPC.}
    \label{fig:swe_sbi_pipeline}
\end{figure}

We formulate personalisation as Bayesian inference of the active parameter vector \(\boldsymbol{\theta}\), conditioned on SWE-derived stiffness dynamics and subject-specific context. Instead of estimating a single best-fit vector, we infer a posterior distribution to quantify uncertainty, parameter compensation, and practical identifiability. Stiffness curves are summarized by curve-derived features, which are combined with subject-specific context to form feature-context training samples. These samples are pooled across subjects to train an amortized inference model that yields subject-specific posterior distributions (Fig.~\ref{fig:swe_sbi_pipeline}).

\paragraph{Feature-based conditioning.}
Simulated and observed stiffness curves were encoded using a hybrid summary representation combining interpretable scalar descriptors with a coarse resampling of the aligned trajectory. This retained mechanically relevant shape information while reducing sensitivity to pointwise noise, temporal misalignment, and redundant samples. Preliminary ablations showed that scalar-only summaries degraded posterior predictive performance. The summary operator \(S(\cdot)\) maps each stiffness curve to \(\mathbf{x}=S(E)\in\mathbb{R}^{46}\), containing 16 scalar descriptors and 30 uniformly resampled trajectory values. The scalar descriptors capture baseline and peak stiffness, amplitude, area under the curve, systolic rise, peak timing, and post-peak relaxation, while the resampled values retain the coarse shape of the systolic rise, peak region, and post-peak relaxation. For case \(d\), and for the \(n\)-th parameter sample
\(\boldsymbol{\theta}_{d,n}\), \(n=1,\ldots,N_{\theta}\), the simulated and
observed feature vectors are
\begin{equation}
    \mathbf{x}_{d,n}
    =
    S\!\left(E_{\mathrm{sim},d}(s;\boldsymbol{\theta}_{d,n})\right),
    \qquad
    \mathbf{x}_{\mathrm{obs},d}
    =
    S\!\left(E_{\mathrm{obs},d}(s)\right).
    \label{eq:feature_vectors}
\end{equation}

\paragraph{Pooled simulation dataset and observation model.}
For each case \(d\), parameter samples are drawn from the common prior, \(\boldsymbol{\theta}_{d,n}\sim p(\boldsymbol{\theta})\), and passed through the context-adapted simulator. The resulting curves are converted into feature vectors \(\mathbf{x}_{d,n}\) as in Eq.~\ref{eq:feature_vectors}. To model observation uncertainty in feature space, each simulated feature vector is augmented with additive Gaussian noise:
\begin{equation}
    \mathbf{x}_{d,n,a}^{\epsilon}
    =
    \mathbf{x}_{d,n}
    +
    \boldsymbol{\epsilon}_{d,n,a},
    \qquad
    \boldsymbol{\epsilon}_{d,n,a}
    \sim
    \mathcal{N}(\mathbf{0},\boldsymbol{\Sigma}_x),
    \label{eq:feature_noise_model}
\end{equation}
where \(a=1,\ldots,N_{\mathrm{aug}}\) indexes feature-noise augmentations. 

In this work, \(\boldsymbol{\Sigma}_x\) is diagonal and non-zero only for maximum stiffness, AUC, post-peak AUC, post-peak decay slope, and time-to-peak, with standard deviations \((1.0,0.5,0.5,5.0,0.03)\) in raw feature units. These scales were selected \textit{a priori} from feature magnitudes, kept fixed across subjects, and used as regularising observation hyperparameters. The resampled trajectory values were not independently perturbed, since they were included as coarse shape descriptors rather than pointwise observations. 

After augmentation, samples from all cases are concatenated into a single pooled training set. Feature and context components are standardized using pooled training-set statistics, and the same transformations are applied to the observed inputs. Each training sample is represented by the conditioning vector
\begin{equation}
    \mathbf{z}_{d,n,a}
    =
    \left[
    \tilde{\mathbf{x}}_{d,n,a}^{\epsilon},
    \tilde{\mathbf{c}}_d
    \right],
    \qquad
    \mathbf{z}_{\mathrm{obs},d}
    =
    \left[
    \tilde{\mathbf{x}}_{\mathrm{obs},d},
    \tilde{\mathbf{c}}_d
    \right],
    \label{eq:conditioning_vectors}
\end{equation}
where brackets denote concatenation and tildes denote pooled standardization. Including \(\mathbf{c}_d\) preserves subject-specific anatomical, physiological, and demographic
information within the pooled training set, while \(\mathbf{z}_{\mathrm{obs},d}\) is standardized but not noise-augmented. The final pooled training dataset is
\begin{equation}
    \mathcal{D}_{\mathrm{pool}}
    =
    \left\{
    \left(
    \boldsymbol{\theta}_{d,n},
    \mathbf{z}_{d,n,a}
    \right)
    :
    d=1,\ldots,D,\;
    n=1,\ldots,N_{\theta},\;
    a=1,\ldots,N_{\mathrm{aug}}
    \right\}.
    \label{eq:pooled_training_dataset}
\end{equation}

\paragraph{Neural posterior estimation.}
We use neural posterior estimation (NPE) to approximate \(p(\boldsymbol{\theta}\mid\mathbf{z})\) from the pooled simulation dataset \(\mathcal{D}_{\mathrm{pool}}\). The estimator \(q_{\eta}(\boldsymbol{\theta}\mid\mathbf{z})\) was implemented with the NPE-C/SNPE-C routine of the \texttt{sbi} toolbox~\cite{Tejero-CanteroEtAl2020,BoeltsEtAl2025}. The density estimator was the default \texttt{sbi} v0.26.1 conditional masked autoregressive flow, using 5 transforms, 50 hidden features, 2 autoregressive blocks, batch size 128, and standardized parameters and conditioning inputs. The estimator is trained by maximizing the conditional log likelihood
\begin{equation}
    \eta^\ast
    =
    \arg\max_{\eta}
    \frac{1}{|\mathcal{D}_{\mathrm{pool}}|}
    \sum_{(\boldsymbol{\theta},\mathbf{z})\in\mathcal{D}_{\mathrm{pool}}}
    \log q_{\eta}(\boldsymbol{\theta}\mid\mathbf{z}).
    \label{eq:npe_objective}
\end{equation}
For each case \(d\), posterior samples are then drawn from \(q_{\eta^\ast}(\boldsymbol{\theta}\mid\mathbf{z}_{\mathrm{obs},d})\) and propagated through the forward model to obtain posterior predictive curves.

\subsection{Experimental setup}
\label{sec:experimental_setup}

All experiments used a single frozen pooled SBI configuration across all analysed subjects, including the prior \(p(\boldsymbol{\theta})\), summary operator \(S(\cdot)\), feature-Gaussian observation model, pooled standardization, NPE architecture, and training protocol. The active-parameter prior was selected to span broad physiologically plausible ranges, informed by literature-reported values~\cite{BanusEtAl2021,MarchesseauEtAl2013a}, previous model calibrations, and prior-support diagnostics. The prior was an independent bounded log-Gaussian prior, log-centred at \(k_0=45\) kPa, \(k_{\mathrm{ATP}}=5\) s\(^{-1}\), and \(k_{\mathrm{SR}}=30\) s\(^{-1}\), with log-space standard deviations \((0.5,0.6,0.45)\) and hard bounds \(k_0\in[10,200]\) kPa, \(k_{\mathrm{ATP}}\in[2,30]\) s\(^{-1}\), and \(k_{\mathrm{SR}}\in[2,60]\) s\(^{-1}\). 

For each evaluated subject, \(N_{\theta}=1000\) successful simulations were generated from the prior using the corresponding subject-specific context. With \(D=6\) evaluated subjects, this yielded \(6000\) forward simulations. Each simulated feature vector was augmented \(N_{\mathrm{aug}}=20\) times using the feature-Gaussian observation model, resulting in \(D \times N_{\theta} \times N_{\mathrm{aug}} = 120{,}000\) pooled training samples. Prior-support diagnostics used to define the retained quantitative analysis set are described in Sec.~\ref{sec:evaluation}. The NPE was trained with a batch size of \(128\) for a maximum of \(100\) epochs, using the standard \texttt{sbi} training procedure~\cite{BoeltsEtAl2025}. Implementation used the \texttt{sbi} toolbox with PyTorch and CUDA acceleration.

\subsection{Evaluation}
\label{sec:evaluation}
Evaluation was performed separately for each subject. Prior support required $\max_j|z_{\mathrm{obs},d,j}|\leq3$, where $j$ indexes the components of the standardized feature-context conditioning vector, and at least 1000 posterior samples within the prescribed prior bounds from $5\times10^5$ candidate draws. The first criterion provides a simple out-of-distribution check, while the second identifies cases with negligible posterior mass within the prior bounds. Cases passing both criteria were considered prior-supported and retained for quantitative posterior analysis. Cases failing either criterion were excluded from quantitative posterior summaries but retained for assessing prior support and model adequacy.

For retained cases, PPCs were computed by sampling \(\boldsymbol{\theta} \sim q_{\eta^\ast}(\boldsymbol{\theta}\mid\mathbf{z}_{\mathrm{obs},d})\) and propagating each sample through the forward simulator to obtain \(E_{\mathrm{sim},d}(s;\boldsymbol{\theta})\) (Eq.~\ref{eq:total_swe_sim}). PPCs assess whether the posterior can reproduce the observed SWE target under the assumed simulator and observation model, not whether the underlying active model parameters are uniquely identified. Predictive agreement was quantified from the posterior predictive median using RMSE, NRMSE, $R^2$, and post-peak RMSE, the latter targeting relaxation dynamics relevant to $k_{\mathrm{SR}}$.

Posterior predictive uncertainty was summarized using the \(90\%\) predictive interval (PI) and its coverage, defined as the fraction of observed cardiac phases lying within the interval. Parameter uncertainty was summarized using posterior medians, \(90\%\) credible intervals (CI), and pairwise posterior dependencies.

As an optimisation-based reference, we performed subject-wise fitting with CMA-ES~\cite{Hansen2007}. For each subject, CMA-ES used the same subject-specific context, passive baseline, and active-parameter bounds as the SBI pipeline, and minimized a curve-level objective combining normalized MAE, correlation mismatch, and peak mismatch. The lowest-loss solution was selected over 10 seeded runs (seeds 200--209; population size 12; 80 generations; maximum 9600 simulator evaluations per subject) and used as a diagnostic point reference, not as ground truth.


\section{Results and Discussion}

\paragraph{Prior-support diagnostics and posterior predictive performance.}
Six subjects were evaluated using the same frozen pooled SBI configuration
(Sec.~\ref{sec:experimental_setup}). Four passed both prior-support criteria and
were retained for quantitative posterior analysis. The remaining two were not excluded because of forward-simulation failure; rather, their observations were insufficiently represented by the fixed prior-predictive training distribution and were retained as
prior-support and model-adequacy cases. In one case, the conditioning vector
satisfied the standardized feature criterion
$(\max_j |z_{\mathrm{obs},j}|=2.16)$, but posterior sampling yielded only one
valid in-prior sample from $5\times10^5$ candidate draws, with an unfiltered
$k_{\mathrm{ATP}}$ median of $69.7~\mathrm{s}^{-1}$. In the other case, the
systolic up-slope lay outside the standardized support threshold $(z=3.14)$,
and no valid in-prior sample was obtained; the unfiltered
$k_{\mathrm{ATP}}$ median was $178.6~\mathrm{s}^{-1}$.

\begin{figure}[!t]
    \centering
    \includegraphics[width=\linewidth]{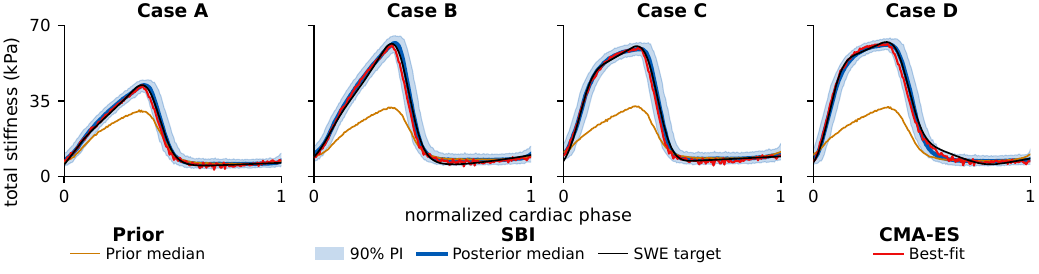}
    \caption{\textbf{Posterior predictive checks.} SBI narrows the prior predictive mismatch of the systolic peak; CMA-ES is shown as a loss-based best-fit reference.}
    \label{fig:ppc_representative_cases}
\end{figure}

Across the four retained cases, the observed values of the five noise-modelled features lay within the posterior predictive 90\% intervals. SBI reduced RMSE from $12.61\pm5.55$ to $1.14\pm0.38\,\mathrm{kPa}$ relative to the prior predictive median, an $89.7\pm4.2\%$ reduction, with high $R^2$ $(0.996\pm0.001)$ (Tab.~\ref{tab:final_ppc_results}). 
This improvement demonstrates posterior-predictive agreement, but does not establish unique identification of the underlying active model parameters. The main posterior correction was improved reproduction of the systolic stiffness peak, which was systematically underestimated by the prior predictive median (Fig.~\ref{fig:ppc_representative_cases}). However, PPCs exposed local adequacy differences beyond global fit metrics: Cases A--C combined accurate medians with near-complete 90\% PI coverage, whereas Case D retained the largest post-peak RMSE and lower coverage, indicating a residual timing or relaxation mismatch.

\begin{table}[ht]
    \centering
    \caption{\textbf{Posterior predictive performance}. RMSEs are in \(\mathrm{kPa}\) and computed against the observed SWE target \(E_{\mathrm{obs},d}(s)\); reduction is relative to the prior predictive median. Coverage refers to the 90\% posterior predictive PI.}
    \label{tab:final_ppc_results}
    \footnotesize
    \setlength{\tabcolsep}{3.5pt}
    \renewcommand{\arraystretch}{1.12}

    \begin{tabular}{@{}lcccccc@{}}
        \toprule
        \textbf{Case} &
        \makecell{\textbf{RMSE}\\\textbf{prior \(\rightarrow\) SBI}} &
        \makecell{\textbf{RMSE}\\\textbf{reduction (\%)}} &
        \textbf{NRMSE} &
        \(R^2\) &
        \makecell{\textbf{Post-peak}\\\textbf{RMSE}} &
        \textbf{Coverage} \\
        \midrule
        \textbf{\small A} & \(4.84 \rightarrow 0.76\) & 84.2 & 0.020 & 0.996 & 0.60 & \(>0.99\) \\
        \textbf{\small B} & \(12.67 \rightarrow 1.40\) & 89.0 & 0.025 & 0.995 & 1.35 & 0.98 \\
        \textbf{\small C} & \(15.33 \rightarrow 0.87\) & 94.3 & 0.016 & 0.998 & 0.76 & 0.99 \\
        \textbf{\small D} & \(17.59 \rightarrow 1.53\) & 91.3 & 0.027 & 0.995 & 1.77 & 0.89 \\
        \bottomrule
    \end{tabular}
\end{table}

\begin{figure}[!t]
    \centering
    \includegraphics[width=\linewidth]{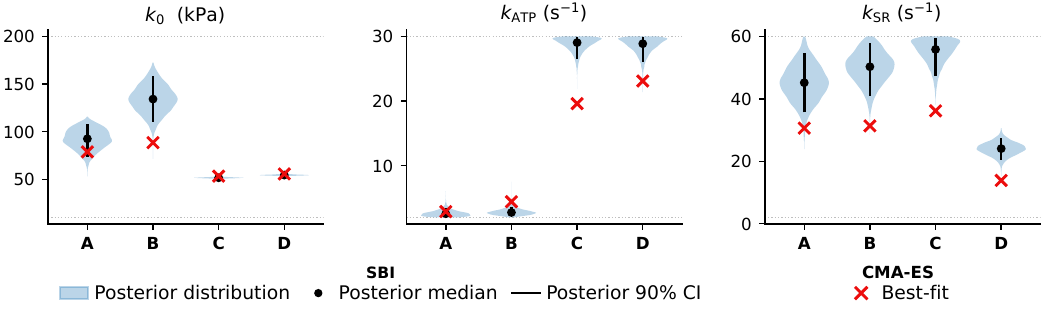}
    \caption{\textbf{Posterior parameter uncertainty}. Distributions, medians, and 90\% CI are obtained from the SBI posterior; crosses indicate CMA-ES best-fit reference estimates.}
    \label{fig:posterior_uncertainty_cases}
\end{figure}

\paragraph{Posterior uncertainty and identifiability.}
Posterior uncertainty was parameter and subject dependent (Fig.~\ref{fig:posterior_uncertainty_cases}). Within the assumed prior, feature set, and forward model, $k_0$ was generally well constrained, consistent with the amplitude information in $E_{\mathrm{obs},d}(s)$. In contrast, $k_{\mathrm{ATP}}$ showed limited practical identifiability: although its posterior could be narrow, wider-prior experiments shifted the inferred contraction-rate distributions. Pairwise posteriors supported this interpretation, with a strong negative $k_0$--$k_{\mathrm{ATP}}$ correlation ($r=-0.84\pm0.15$), indicating compensation between active-stiffness amplitude and contraction rate. Correlations involving $k_{\mathrm{SR}}$ were weak on average ($r(k_0,k_{\mathrm{SR}})=0.09\pm0.12$, $r(k_{\mathrm{ATP}},k_{\mathrm{SR}})=-0.03\pm0.15$), suggesting weaker and subject-dependent relaxation-rate information. Consistently, a preliminary synthetic recovery check ($N=400$) showed near-nominal 90\% coverage for $k_0$ and $k_{\mathrm{ATP}}$ (92.5\%, 90.5\%) but marked under-coverage for $k_{\mathrm{SR}}$ (26.5\%). Together with the excluded cases and the degraded coverage observed when widening the timing-related prior bounds, these results motivate explicit prior-support diagnostics, improved prior design, and future simulation-based calibration. 

These results describe practical identifiability conditional on the selected
prior, observation model, fixed passive component, and $0$D forward model.
Spherical symmetry, isotropy, and fixed circulatory parameters neglect spatial
heterogeneity, fibre architecture, and subject-specific loading. The observed
parameter dependencies may therefore reflect both limited SWE information and
structural model mismatch. 

\paragraph{Diagnostic comparison with CMA-ES.}
CMA-ES was used as an optimisation-based point reference following the protocol
described in Sec.~\ref{sec:evaluation}. Across the four prior-supported cases,
SBI posterior predictive medians achieved RMSEs of 0.76, 1.40, 0.87, and
$1.53~\mathrm{kPa}$, compared with 1.36, 2.15, 2.06, and
$2.02~\mathrm{kPa}$ for the CMA-ES best-fit estimates in Cases A--D.
This comparison is diagnostic rather than competitive, since SBI summarizes a
posterior predictive distribution, whereas CMA-ES returns a single
loss-minimizing parameter vector. Differences may therefore reflect parameter
compensation, sensitivity to the optimisation objective, or residual model
mismatch, and should not be interpreted as evidence of unique parameter
recovery.


\section{Conclusion}

We presented a context-conditioned SBI framework for probabilistic personalisation of active myocardial mechanics from SWE-derived stiffness dynamics. Among six healthy volunteers, four met the prior-support criteria and showed improved posterior-predictive agreement, whereas two revealed inadequate representation within the prior-predictive distribution. Posterior analysis identified $k_0$--$k_{\mathrm{ATP}}$ compensation, parameter-specific uncertainty, and weaker constraint of $k_{\mathrm{SR}}$. Predictive agreement does not establish unique recovery of the underlying model parameters, and the reported uncertainty remains conditional on the selected prior, prescribed feature-noise covariance, precomputed passive component, fixed circulatory parameters, and simplified $0$D model. The small healthy-volunteer sample limits assessment of between-subject variability and generalisability to pathological populations. Nevertheless, these preliminary results support SBI as a means of distinguishing parameters constrained by SWE from those that remain ambiguous, thereby reducing over-interpretation of a single fitted solution. Clinical translation will require observation uncertainty calibrated from repeated acquisitions, improved subject-specific modelling of passive mechanics and loading, and validation in larger pathological cohorts.

\subsubsection*{Acknowledgments.} This work was supported by ANR through PEPR Digital Health ChroniCardio (22-PESN-0015), 3IA C\^ote d'Azur/IA Cluster (ANR-19-3IA-0002, ANR-23-IACL-0001), and IHU Liryc (ANR-10-IAHU-04); by France 2030 and Next Generation EU through MediTwin; and by the ERC Horizon Europe 5D ULTRAFAST HCM project (101220327). We acknowledge
the OPAL infrastructure (Universit\'e C\^ote d'Azur) for computational support. Finally, we thank Giulio Corallo for code assistance and manuscript feedback, and
Giuseppe Orlando for manuscript feedback and helpful discussions.
\subsubsection*{Disclosure of Interests.} The authors have no competing interests to declare that are relevant to the content of this article.

%
%

\bibliographystyle{splncs04}
\bibliography{bibliography}

\end{document}